\documentclass[12pt]{article}

\newcommand{\be}{\beta}
\newcommand{\ga}{\gamma}
\begin{document}

\title{\LARGE Series solution of the $1+2$ continuous Toda chain }
\author{ D.B  Fairlie$\footnote{e-mail david.fairlie@durham.ac.uk}$
\thanks{Durham University, Durham, England.},\  A.N.Leznov $\footnote{e-mail:
andrey@buzon.uaem.mx}$\thanks{ Universidad Autonoma del Estado de Morelos,
CIICAp,Cuernavaca, Mexico}\ and R.Torres-Cordoba} \date{}
\maketitle
\begin{abstract}
A way to obtain the series solutions of the $1+2$ dimensional
continuous Toda chain is presented.
\end{abstract}
 
\section{Introduction}
The  usual form of the equation under consideration is the following one;
\begin{equation}
 \rho_{y,x}=(e^{\rho})_{z,z},\quad (\ln u)_{y,x}=u_{z,z},\quad u \equiv e^{\rho} \label{1}
\end{equation}
Here $\rho(x,y,z)$ is an  unknown function of three independent
variables. This equation arises as a reduction of the Plebansky
equation \cite{1} describing self-dual four dimensional $(0+4),(2+2)$ gravity.
In this connection it was considered in \cite{2}. Equation (\ref{1})
can also be obtained  as a limit of the discrete Toda chain
\[
\rho_{y,x}=e^{\rho_{n+1}}-2e^{\rho_n}+e^{\rho_{n-1}}
\]
under appropriate rescaling   $(n\to z)$. This is the reason for the title of
the present paper.
A solution of the two dimensional reduction of (1) (\ref{1}) ($\rho=\rho
(z,y+x)$ was found in implicit form in \cite{3}. Infinite series
solutions of the symmetry equation corresponding to (\ref{1}) were
found in \cite{4}. But the connection between series solutions of the symmetry
equation with  the solution of the initial system (\ref{1}) has not been found. But
general theory gives a guarantee that with each solution of symmetry equation is
connected with an analytical solution of the initial system in explicit or
implicit form.
The goal of the present paper is fill this gap and demonstrate a way how an
analytical solution of
(\ref{1}) is connected with the solution  constructed in \cite{4} of the
symmetry equation.
In \cite{LP}, the Plebansky equation was represented in the form of two equations of
the first order for two unknown functions. One of which satisfies the Plebansky
equation by itself, the second one satisfies the corresponding   symmetry
equation. It is possible to find by an independent way the solution of symmetry
equation  a in recurrence form. As  was remarked the
above equation under consideration in the present paper is a reduction of
Plebansky equation and so it is possible to try to solve it by the same methods.
 
\section{Preliminary manipulations. Short excursion into \cite{4}  }
Let us rewrite (\ref{1}) in the form of a system of two equations of the first
degree
\begin{equation}
(\ln u)_y=T_z,\quad  u_z=T_x,\quad  (\ln u)_{x,y}=u_{zz},\quad \left(\frac{T_x}{u}\right)_y=T_{zz} \label{2}
\end{equation}
or as the initial equation is symmetrical with respect to exchange of the
variables $x,y$, the following  is also a possible form;
\begin{equation}
(\ln u)_x=w_z,\  u_z=w_y,\quad  (\ln u)_{x,y}=u_{zz},\ \left(\frac{w_y}{u}\right)_x=w_{zz}\label{3}
\end{equation}
The symmetry equation arises from the initial one after differentiation by  some
arbitrary parameter and  considering this derivative as a new unknown function.
In the case under consideration this equation is;
\[
\dot u=S,\ \left(\frac{S}{ u}\right)_{x,y}=S_{z,z},\ S=T_x (w_y),\ \left(\frac{T_x}{
u}\right)_y=T_{zz},\quad \left(\frac{w_y}{ u}\right)_x=w_{zz}
\]
Thus we see that the linear system of equations of the first few lines of this
section is connected with the symmetry equation  of the initial system.
In \cite{4} we have obtained series solutions of the symmetry equation in
integro-differential terms of the function $u$. And thus it is possible use these
expressions in the system $T,u$ and obtain two self consistent equation instead
of only one equation for the function $u$. It is obvious that in this way we will
not be able to obtain a general solution for the equation for $u$  but only its
partial soliton like series solutions. Resolving the second equation of
(\ref{2}) $u=\theta_x,T=\theta_z$ we rewrite (\ref{1}) in the form
\[
\theta_{y,x}=\theta_x \theta_{z,z},\quad w_{y,x}=w_y w_{z,z} \label{4}
\]
In \cite{4} it was shown that solution of symmetry equation $T$ may be obtained
in terms of $r$ $\alpha_n$ functions which satisfy the the following recurrence
relations
\[
\alpha_{n-1}=\int dy \frac{(u^{n+1} \alpha_n)_z}{ (n+1)u^{n+1}},\quad \frac{{(u^{n+1}
\alpha_n)}_x}{ u^{n+1}}=(n+1) {\alpha_{n-1}}_z
\]
Eliminating  $\alpha_{n-1}$ from both equations we arrive at the equation for
$\alpha_n$ function in the form
\[
\left(\frac{{(u^{n+1} \alpha_n)}_z}{ u^n}\right)_z=\left(\frac{(u^{n+1}{ \alpha_n)}_x}{
u^{n+1}}\right)_y,\quad (u^{n+2} \alpha^n_z)_z=(u^{n+1} \alpha^n_y)_x
\]
The left and right equations are the same. From these expressions it follows that
there exists an obvious solution $\alpha^n=1$ which leads to  a finite solution
for $T$.
The solution for $T$ becomes
\[
T=u\alpha^0=u\int dy \frac{(u^2 \alpha^1)_z}{ 2u},\quad \theta_y-\frac{\theta^2_z}{
2}=\frac{u^2 \alpha_1}{ 2}
\]
The second equality is obtained from the first one after the substitution
$T=\theta_z,u=\theta_x$ and differentiation of the subsequent  $\frac{\theta_z}{
\theta_x}$ with respect to the  argument $y$ and integration once over $z$.
 
\section{Generalization of R.Ward's solution.}
This section explains why analytic solutions of Ward exist at all.
The simplest solution of the symmetry equation is a linear combination  of
derivatives of the  functions $u$ $S=w_y=u_z=au_x+bu_y+cu_z$
The solution of Richard Ward corresponds to choosing $c=1,a=-b$. In the case  $c$
is not equal to zero we have $u_z=Au_x+Bu_y$ and the second system under this
additional condition becomes
\[
u_x=u w_z,\quad u_z=w_y,\quad u_x=u (A w_x+B w_y),\quad A u_x+B u_y=w_y
\]
Let us seek  a solution of this system in the form
\[
x=\theta(u,w),\quad y=\sigma(u,w)
\]
The system of equations defining derivatives of $u,w)$ with respect
to space coordinates $(x,y)$ is the following one;
\[
1=\theta_u u_x+\theta_w w_x,\quad  0=\sigma_u  u_x+\sigma_w w_x
\]
\[
0=\theta_u u_y+\theta_w w_y,\quad  1=\sigma_u  u_y+\sigma_w w_y.
\]
After solving  the last system,
\[
u_x=\frac{\sigma_w}{ D},\quad w_x=-\frac{\sigma_u}{ D},\quad
u_y=-\frac{\theta_w}{ D},\quad w_y=\frac{\theta_u}{ D}
\]
and substitution into the previous one we arrive at a linear system of equations
for $\theta(u,w)$ and $\sigma(u,w)$.
\[
\sigma_w=u(-A \sigma_u+B \theta_u),\quad   A \sigma_w-B \theta_w=\theta_u
\]
The last system after eliminating (for instance) the function $\sigma$ leads to
an equation of the second order with separable variables
\[
u \theta_{u,u}+\frac{1}{ A}\theta_{w,u}+\frac{B}{ A}\theta_{w,w}=0
\]
The case considered by Ward corresponds to the limiting case $A\to \infty,\  B\to
\infty,\ \frac{B}{ A}\to -1$.
\section{The zero order term of series solution to the symmetry equation }
In the case $\alpha_0=1$ from the general formula it follows that $T=u$ or $w=u$
and from the corresponding formulas of the previous section we obtain $u_x=u_z$ or
$u_y=u_z$. These are particular cases of the generalized Ward construction of
the previous section. The first equations in this cases lead $u_y=uu_z$. This
well the known Monge equation (the equation of Hamilton-Jacobi for free motion in one
dimension) with general solution $ z+y+u x=F(u)$ or $ z+x+u y=F(u)$. It is not
difficult to connect these solutions with the generalized Ward solution of the previous
section.

\section{The first term of  the  symmetry equation series solution}
In the case $\alpha_1=1$ in connection with the recurrence procedure we obtain 
$\alpha_0=\int d yu_z$ and solution for $T$ takes the form
\[
T=u(\int d y u_z),\quad U=\int d y u,\quad u=U_y ,\quad T=U_y U_z
\]
and the equations which are necessary to resolve are the following;
\[
U_{y,z}=(U_y U_z)_x ,\quad (\ln U_y)_y=(U_y U_z)_z,\quad (\ln U_y)_x=U_{z,z}
\]
Taking into account the last equation the first one leads  to a relation between the
derivatives (after  integration once with respect to the  argument $z$) in the form
$\ln U_y=\frac{1}{ 2}U^2_z+U_x,\  U_y=e^{\frac{1}{ 2}U^2_z+U_x}$.
Let us seek a solution of these equations using the following parametrisation
\[
x=X(U_x,U_z,y),\quad z=Z(U_x,U_z,y),\quad U_{x,x}=\frac{Z_\ga}{ D} \quad
U_{x,z}=-\frac{Z_\be}{ D}= -\frac{X_\ga}{ D}\quad U_{z,z}=\frac{X_\be}{ D}
\]
where  indices   $\be,\ga$ denote respectively the first and second arguments  of the functions
$X,Z$. $X=W_\be,Z=W_\ga$ and equation transorm to a linear equation of second order with
separable variables
\[
-2 W_{\be,\ga}+W_{\ga,\ga}=W_{1,1},\quad W=e^{k \be} U(\ga),\quad -2 k U_\ga+ U_{\ga,\ga}=k^2 U
\]
 It is possible obtain the dependence of the function $W$ after solution of two
equations which arise after differentiation of the previous equations by the 
argument $y$
\[
X_\be u_{x,y}+X_\ga u_{z,y}+X_y=0,\quad Z_\be u_{x,y}+Z_\ga u_{z,y}+Z_y=0,
\]
 remembering that $U_y=e^{\frac{1}{2}U^2_z+U_x}$. After trivial manipulations
we obtain for $W$
\[
W=-ye^{\frac{1}{2}U^2_z+U_x}+W^L,\quad x=W_\be=ye^{\frac{1}{ 2}U^2_z+U_x}+W^L_{\be}\quad
x=W_1=y 2 e^{\frac{1}{2}U^2_z+U_x}+W^L_\be
\]
where $W^L$ is solution of the linear equation obtained above. Solution of Toda
chain of the begining of this paper is given by connection $u=U_y=e^{\frac{1}{
2}U^2_z+U_x}$.
\subsection{One simple example}
It is easy to check that $W^L= 2 e^{-\be}$ is an explicit solution of the linear
equation and thus $ W=ye^{\frac{1}{ 2}U^2_z+U_x}+2 e^{-\be}$.
The implicit form of the solution is given by
\[
x=W_\be=-ye^{\frac{1}{2}U^2_z+U_x}-2e^{-\be},\quad z=W_\ga=-y 2 e^{\frac{1}{
2}U^2_z+U_x}+e^{-\be},\quad U_z=\ga,U_x=\be
\]
From these expressions immediately follows the equation $ U_{z,z}+2U_z
U_{z,x}-U_{x,x}=0$, which is equivalent to our linear system above.
With the help of this equation it is not difficult to check that $ T= U_y U_z=2
e^{\frac{1}{ 2}U^2_z+U_x},\quad U_y=e^{\frac{1}{2}U^2_z+U_x}$
satisfy the equations
\[
U_{y,z}=T_x,\quad (\ln U_y)_y=T_z
\]
and thus $u=U_y$ is a solution of the equation of title of this paper.
More intriguing and interesting are the following comments.
We rewrite equations which define an implicit solution in a form $u=U_y=e^{\frac{1}{
2}U^2_z+U_x}\equiv e^{\frac{1}{2}\ga^2+1}$ in equivalent form
\[
x=-y u-2\frac{1}{ u}e^{\frac{1}{2}\ga^2},\quad z=-2 y u+\frac{1}{u}e^{\frac{1}{2}\ga^2}
\]
After excluding terms containing $y$ we arrive at a quadratic equation for
determining the variable $\ga$
\[
\ga^2+\frac{z}{ yu}\ga+\frac{v}{ yu}+1=0
\]
Substituting the solution of this equation into the first or second equations we
come to equation determining in implicit form the function $u$.
It is obvious that  to obtain this equation is not a very
simple problem.
In Appendix we present an alternative method of solution of the problem of this
section.

\section{Second step}

In the case where $\alpha_2=1$ in connection with the recurrence procedure we obtain 
$\alpha_1=\int d yu_z=U_z$ ,$\alpha_0=\int d y \frac{u^2\alpha_1}{2 u}=\int d
y ( U_{y,z}U_z+\frac{1}{2}U_yU_{z,z})=\frac{1}{2}(U^2_z+U_x)$ and the solution for
$T$ takes the form
\begin{equation}
T=\frac{1}{2}U_y (U^2_z+U_x),\quad U=\int d y u,\quad u=U_y
\end{equation}
The equations which are necessary to resolve are the following ones;
\[
U_{y,z}=(T)_x=\frac{1}{2}[U_{y,x} (U^2_z+U_x)+2U_zU_{z,x}+U_{x,x}] ,\quad
(\ln U_y)_y=(T)_z,\quad  U_{y,x}=U_y U_{z,z}
\]
Let us introduce notations
\[
U_y=e^c, \quad U_x=a , \quad U_z=b ,\quad \alpha=\frac{1}{2}(b^2+a)
\]
Equations above together with introduced notations lead to the following
system of equations
\[
\pmatrix{ a \cr
         b \cr
          c \cr}_z=\pmatrix{ 0 & 1 & 0 \cr
                             0 & 0 & 1 \cr
                             \frac{1}{2} & b & \alpha \cr}\pmatrix{ a \cr
                                                                    b \cr
                                                                  c
\cr}_x\equiv L\pmatrix{ a \cr
                                                                    b \cr
                                                                  c
\cr}_x,\quad
\pmatrix{ a \cr
          b \cr
          c \cr}_y=e^c L\pmatrix{ a \cr
                                                                    b \cr
                                                                    c \cr}_z
\]
As in the cases above we will seek solution of these equation by implicit
substitution
\[
x=X(a,b,c),\quad,z=Z(a,b,c),\quad y=Y(a,b,c)
\]
After differentiation all these equalities with respect to independent
arguments of problem and introduction matrix
$ V=\pmatrix{ X_a & X_b & X_c \cr
              Z_a & Z_b & Z_c \cr
             Y_a & Y_b & Y_c \cr}$ we have
\[
\pmatrix{ a \cr
          b \cr
          c \cr}_x=V^{-1}\pmatrix{ 1 \cr
                                   0 \cr
                                   0 \cr},\quad \pmatrix{ a \cr
          b \cr
c \cr}_z=V^{-1}\pmatrix{ 0 \cr
                                   1 \cr
                                   0 \cr}    ,\quad \pmatrix{ a \cr
          b \cr
          c \cr}_y=V^{-1}\pmatrix{ 0 \cr
                                   0 \cr
                                   1 \cr}
\]
Substituting these expressions in linear system equations of the first order
we obtain
\[
 \pmatrix{ 0 \cr
            1 \cr
            0 \cr}=VLV^{-1}\pmatrix{ 1 \cr
                                     0 \cr
                                     0 \cr},\quad \pmatrix{ 0 \cr
                                   0 \cr
                                   1 \cr}=e^c VLV^{-1}\pmatrix{ 0 \cr
                                   1 \cr
                                   0 \cr}
  \]
 The last equations allow to reconstruct explicit form matrix $VLV^{-1}$
\[
VLV^{-1}=\pmatrix{ 0 & 0 & \frac{1}{2}e^c \cr
                   1 & 0 & be^c \cr
                   0 & e^{-c} & \alpha \cr}=\pmatrix{ 1 & 0 & 0 \cr
                                                     0 & 1 & 0 \cr
                                                     0 & 0 &
e^{-c}\cr}L^T\pmatrix{ 1 & 0 & 0 \cr
                                                     0 & 1 & 0 \cr
                                                     0 & 0 & e^c\cr}
 \]
The first two columns are direct consequence of equations above. The last
column arises from the fact $Trace  (VLV^{-1})^n=Trace L^n$.
Now we come to linear system of equations for determining $X,Z,Y$ functions
\[
\pmatrix{ 1 & 0 & 0 \cr
          0 & 1 & 0 \cr
          0 & 0 & e^c\cr} VL= L^T\pmatrix{ 1 & 0 & 0 \cr
                                                     0 & 1 & 0 \cr
                                                     0 & 0 & e^c\cr}V
\]
\[
\pmatrix{ 1 & 0 & 0 \cr
          0 & 1 & 0 \cr
          0 & 0 & e^c\cr} V=\pmatrix{ X_a & X_b & X_c \cr
              Z_a & Z_b & Z_c \cr
             (e^cY)_a & (e^cY)_b & (e^cY)_c-e^cY \cr}
\]
The first two columns are direct consequence of equations above. The last
column arises from the fact $Trace  (VLV^{-1})^n=Trace L^n$.
Now we come to linear system of equations for determining $X,Z,Y$ functions
\[
\pmatrix{ 1 & 0 & 0 \cr
          0 & 1 & 0 \cr
          0 & 0 & e^c\cr} VL= L^T\pmatrix{ 1 & 0 & 0 \cr
                                                     0 & 1 & 0 \cr
                                                     0 & 0 & e^c\cr}V
\]
\[
\pmatrix{ 1 & 0 & 0 \cr
          0 & 1 & 0 \cr
          0 & 0 & e^c\cr} V=\pmatrix{ X_a & X_b & X_c \cr
              Z_a & Z_b & Z_c \cr
             (e^cY)_a & (e^cY)_b & (e^cY)_c-e^cY \cr}
\]
$e^cY$ we will denote by $Y$. System of 9 equations are the following one
\[
\pmatrix{ \frac{1}{2}(Y_a-X_c) & \frac{1}{2}Y_b-X_a-b X_c & {1\over
2}(Y_c-Y)-X_b- \alpha X_c \cr
     X_a+bY_a-\frac{1}{2}Z_c & X_b+bY_b-Z_a-b Z_c  & X_c+b(Y_c-Y)-Z_b- \alpha
Z_c \cr
Z_a+ \alpha Y_a -\frac{1}{2}(Y_c-Y) & Z_b+ \alpha Y_b-Y_a-b(Y_c-Y) & Z_c-Y_b
\cr}=0
\]
Elements $M_{1,1}$ and $M_{3,3}$ lead to a parametrization $ X=R_a, Y=R_c,
Z=R_b+f(a,b)$. Elements $M_{2,1}$ and  $M_{1,2}$ both lead to equation $
(R_a+bR_c)_a=\frac{1}{2}R_{b,c}$.  Element  $M_{2,2}$ allow to conclude
function $f$ depend only from one argument $b$.
Elements $M_{3,1}$ and  $M_{1,3}$ are the sane and lead to equation $
(R_b+\alpha R_c)_a=\frac{1}{2}R_{c,c}$ . And at last elements $M_{3,2}$ and 
$M_{2,3}$ pass to a third equation in the form $ (R_b+f(b)+\alpha
R_c)_b=(R_a+bR_c)_c$.
Thus we have three equations which it is necessary to resolve
\[
 R_a+bR_c)_a=\frac{1}{2}R_{b,c},\quad (R_b+\alpha R_c)_a={1\over
2}R_{c,c},\quad (R_b+f(b)+\alpha R_c)_b=(R_a+bR_c)_c
 \]
or
\[
R_a+bR_c=W_b,\quad \frac{1}{2}R_c=W_a,\quad R_b+f(b)+\alpha R_c=W_c
\]
 For further calculations it will be more suitable Variables
$b,\alpha=\frac{1}{2}(a+b),c$. In these variables the system equations above
looks as
\[
R_c=W_{\alpha},\quad \frac{1}{2}R_{\alpha}+bR_c=W_b+bW_{\alpha},\quad
R_b+bR_{\alpha}+\alpha R_c=W_c
\]
First equation give $R=Q_{\alpha}, W=Q_c$ Substituting both others equation
we pass to system of two equations
 \[
 Q_{\alpha,\alpha}=2Q_{b,c},\quad Q_{\alpha,b}+b Q_{\alpha,\alpha}+\alpha
Q_{\alpha,c}=Q_{c,c}
  \]
Let us seek solution of this linear system above in Laplace-Furier forma
 \[
 Q=\int d k d p e^{k \alpha+p b+{k^2\over 2p}c} f(p,k)
  \]
 The first equation is satisfied automatically. The second one leads to a
differential equation of the first order in partial derivatives for the
determination of the under the integral function $f(k,p)$, which for the function
$F\equiv k^3 f$ is
\[
2 p \frac{\partial F}{\partial p}+ k\frac{\partial F}{\partial k}=\left(\frac{2
p^2}{ k}-\frac{k^2}{ 2p}\right) F
\]
with the  obvious solution
\begin{equation}
F=k^3 f=e^{\frac{p^2}{3 k}+\frac{1}{2}\ln k \frac{k^2}{ p}} \phi\left(\frac{k^2}{p}\right)
\end{equation}
where $\phi$ is an arbitray function of the argument $\left(\frac{k^2}{p}\right)$.
\section{Outlook}
We have presented a new idea, unknown up to now to the best of our knowledge in theory 
of integrable systems connected with the symmetry equation of the initial system. We have also presented some
non-trivial solutions of the $2+1$ continu Toda chain. 
 
\end{document}